\documentclass[a4paper, twocolumn]{article}

\usepackage[english]{babel}
\usepackage{csquotes}
\usepackage[top=18.5mm, bottom=18.5mm, left=18.5mm, right=18.5mm]{geometry}
\usepackage{bm}
\usepackage[hidelinks]{hyperref}
\usepackage[sorting=none, giveninits=true, maxnames=4]{biblatex}
\AtEveryBibitem{\clearfield{note}}

\usepackage{authblk}

\usepackage{titlesec}
\titleformat{\section}{\raggedright\bfseries}{\thesection}{1em}{\MakeUppercase}
\titleformat{\subsection}{\raggedright\bfseries}{\thesubsection}{1em}{}

\usepackage{caption} 
\usepackage{amsmath}
\usepackage{amssymb}
\usepackage{xfrac}

\usepackage{sistyle}

\newcommand{\norm}[1]{\left\lVert#1\right\rVert}
\newcommand*{\tran}{^\mathrm{T}}
\DeclareMathOperator{\sign}{sign}
\DeclareMathOperator{\MSE}{MSE}
\DeclareMathOperator{\MAE}{MAE}

\title{Stable Differentiable Modal Synthesis\\for Learning Nonlinear Dynamics}
\author[1]{Victor Zheleznov\thanks{To whom correspondence should be addressed. E-mail: v.zheleznov@ed.ac.uk}\thanks{This work was supported by the SGSAH AHRC Doctoral Training Partnership [grant number AH/R012717/1]; and the University of Edinburgh.}}
\author[2]{Stefan Bilbao}
\author[1]{Alec Wright}
\author[3]{Simon King}
\affil[1]{Acoustics and Audio Group, University of Edinburgh, Edinburgh, UK}
\affil[2]{STMS (UMR9912), IRCAM, CNRS, Sorbonne Université, Paris, France}
\affil[3]{Centre for Speech Technology Research, University of Edinburgh, Edinburgh, UK}
\date{}

\begin{document}
\maketitle

%%%%%%%%%% ABSTRACT %%%%%%%%%%
\begin{abstract}
Modal methods are a long-standing approach to physical modelling synthesis. Extensions to nonlinear problems are possible, leading to coupled nonlinear systems of ordinary differential equations. Recent work in scalar auxiliary variable techniques has enabled construction of explicit and stable numerical solvers for such systems. On the other hand, neural ordinary differential equations have been successful in modelling nonlinear systems from data. In this work, we examine how scalar auxiliary variable techniques can be combined with neural ordinary differential equations to yield a stable differentiable model capable of learning nonlinear dynamics. The proposed approach leverages the analytical solution for linear vibration of the system's modes so that physical parameters of a system remain easily accessible after the training without the need for a parameter encoder in the model architecture. Compared to our previous work that used multilayer perceptrons to parametrise nonlinear dynamics, we employ gradient networks that allow an interpretation in terms of a closed-form and non-negative potential required by scalar auxiliary variable techniques. As a proof of concept, we generate synthetic data for the nonlinear transverse vibration of a string and show that the model can be trained to reproduce the nonlinear dynamics of the system. Sound examples are presented.
\end{abstract}

%%%%%%%%%% BODY %%%%%%%%%%
%%%%%%%%%% INTRODUCTION %%%%%%%%%%
\section{INTRODUCTION}

The aim of physical modelling synthesis is to generate sound by numerically solving ordinary or partial differential equations (ODEs/PDEs) that describe the dynamics of an acoustic system. Throughout its long history, various simulation techniques have been employed, including modal synthesis \cite{Morrison1993}, finite-difference time-domain methods \cite{BilbaoNSS} and port-Hamiltonian methods \cite{Lopes2016}. Numerical stability of such approaches has always been a critical topic in the literature, which has seen a resurgence due to the scalar auxiliary variable (SAV) technique \cite{Bilbao2023, Russo2024, Risse2025}. Under the non-negativity condition on potential energy, the SAV technique provides a way to construct explicit and provably stable numerical solvers for general classes of nonlinear systems.

More recently, machine learning approaches have seen an increasing interest in numerical simulation \cite{Chen2018, Raissi2019, Li2021, Lu2021}, including for modelling of distributed musical systems \cite{Parker2022, Martin2023, Diaz2024, Lee2024, Luan2025, Diaz2025JAES}, and can yield differentiable models that are well-suited for learning from data while respecting physical priors \cite{Diaz2025DAFx}. In contrast with classical approaches, machine learning methods usually operate without stability guarantees. Extrapolation in time beyond intervals seen during training is a significant challenge and can lead to a rapid degradation in solution accuracy \cite{Parker2022, Martin2023, Diaz2024}. In addition to stability concerns, a common drawback of machine learning approaches is that sampling rate and physical parameters of a system, affecting pitch, timbre and other sonic characteristics, cannot be changed after training \cite{Parker2022, Martin2023, Diaz2024, Luan2025, Diaz2025JAES}, or the model architecture relies on a parameter encoder to condition the solution \cite{Lee2024}, leading to more trainable parameters and the requirement of a larger dataset containing ground truth data for all desired configurations of a system.

In this paper we expand our physics-informed machine learning framework presented at DAFx25 conference \cite{Zheleznov2025} by introducing stability guarantees. We use modal decomposition to construct a system of finite dimension, separate the linear and nonlinear parts of the problem and replace only a dimensionless memoryless nonlinearity that describes coupling between the modes with a neural network. In contrast to our previous work \cite{Zheleznov2025} that relied on multilayer perceptrons (MLPs), we impose additional architectural constraints \cite{Massi2024} through gradient networks (GradNets) \cite{Chaudhari2025} that allow an interpretation in terms of a closed-form and non-negative potential function. This enables application of the SAV technique for numerical integration of neural ordinary differential equations (NODEs) \cite{Chen2018}, leading to a stable differentiable model that can learn nonlinear dynamics from data. As a result of separating the linear vibration of a system from the neural network, physical parameters remain easily accessible and the model generalises to physical parameters, sampling rates and time scales not seen during training.

The paper is organised as follows. A simple model of nonlinear transverse string vibration is described in Section \ref{sec: string}. Section \ref{sec: modal} derives modal equations for a string, which are then discretised in time using an explicit and stable numerical solver in Section \ref{sec: numerical_solver}. Section \ref{sec: model} outlines the proposed differentiable model based on the obtained modal equations, which is evaluated for the case study of nonlinear transverse string vibration in Section \ref{sec: evaluation}. Sound examples are available on the accompanying page\footnote{\url{https://victorzheleznov.github.io/jaes-modal-node}}.

%%%%%%%%%% NONLINEAR TRANSVERSE STRING %%%%%%%%%%
\section{NONLINEAR TRANSVERSE STRING VIBRATION}
\label{sec: string}

The general equation of motion describing the transverse nonlinear vibration of a string in a single polarisation is:
\begin{equation}
\mathcal{L} u
=
\mathcal{F}
+
\mathcal{F}_{\mathrm{e}}.
\label{eq: string}
\end{equation}
Here $u = u(x, t)\!\!: [0, L] \times \mathbb{R}^+ \rightarrow \mathbb{R}$ denotes the transverse displacement of a string of length $L$ and depends on spatial coordinate $x$ in \SI{}{m} and time $t$ in \SI{}{s}. Initial conditions are assumed to be zero. The string is assumed to be simply supported at both ends, implying the following boundary conditions:
\[
u(0, t)
=
\partial_x^2
u(0, t)
=
u(L, t)
=
\partial_x^2
u(L, t)
=
0,
\;
\forall
t
\in
\mathbb{R}^+
,
\]
where $\partial_x$ represents a partial derivative with respect to $x$. Output is assumed to be drawn directly from the string displacement at position $x_{\mathrm{o}}$ as $w(t) = u(x_{\mathrm{o}}, t)$.

%%%%%%%%%% LINEAR VIBRATION %%%%%%%%%%
\subsection{Linear Vibration}

The linear part of the string vibration is encapsulated in the operator $\mathcal{L}$, defined as:
\[
\mathcal{L}
=
\rho A
\partial_{t}^2
-
T_0 \partial_{x}^2
+
EI \partial_{x}^4
+
2 \sigma_0 
\rho A
\partial_t
-
2 \sigma_1
\rho A
\partial_t\partial_{x}^2,
\]
where $\partial_t$ represents a partial derivative with respect to $t$. Physical parameters that appear in $\mathcal{L}$ are: the material density $\rho$ in \SI{}{kg.m^{-3}}; the string cross-sectional area $A = \pi r^2$ in \SI{}{m^{2}} for a string of radius $r$; the tension $T_0$ in \SI{}{N}; Young's modulus $E$ in \SI{}{N.m^{-2}}; and moment of inertia $I = \frac{1}{4} \pi r^4$ in \SI{}{m^{4}}. Frequency-independent and dependent loss is characterised by parameters $\sigma_0 \geq 0$ and $\sigma_1 \geq 0$, respectively. See \cite{BilbaoNSS} for more on these terms in the context of linear string vibration.

%%%%%%%%%% NONLINEARITY %%%%%%%%%%
\subsection{Nonlinearity}

Nonlinear dynamics of the string are described in a force density $\mathcal{F}$ that can be expressed through the potential $\mathcal{V}$:
\[
\mathcal{F}
(x, t)
=
\frac{EA - T_0}{2}
\partial_x
\big(
\mathcal{V}'(\xi)
\big),
\quad
\xi
=
\xi
(x, t)
\triangleq
\partial_x u,
\]
where prime denotes a derivative of a scalar function.
A general model for the potential $\mathcal{V}$ is given by Morse and Ingard \cite{MorseAcoustics} and includes both longitudinal and transverse motion of a string in two polarisations. In this work we neglect the longitudinal motion and one of the two polarisations, leading to the following nonlinear function \cite{Bilbao2023}:
\begin{equation}
\mathcal{V}
(\xi)
=
\Big(
\sqrt{1 + \xi^2}
-
1
\Big)^2
.
\label{eq: potential}
\end{equation}
The potential \eqref{eq: potential} can be approximated by Taylor series as $\frac{1}{4}\xi^4$ which corresponds to the model used in our previous work \cite{Zheleznov2025}. Further simplification can lead to a Kirchhoff-Carrier model \cite{KirchhoffVorlesungen, Carrier1945}, where the nonlinearity is averaged over the length of the string. Compared to the Kirchhoff-Carrier model, which adequately reproduces only the pitch glide effect, the potential \eqref{eq: potential} is capable of capturing other perceptually important effects such as phantom partials.

%%%%%%%%%% PLUCKING EXCITATION %%%%%%%%%%
\subsection{Plucking Excitation}

The string is excited by a pointwise external force $\mathcal{F}_{\mathrm{e}}$, which can be modelled as:
\[
\mathcal{F}_{\mathrm{e}}
(x, t)
=
\delta(x - x_{\mathrm{e}})
f_{\mathrm{e}}(t)
,
\]
where $\delta(x - x_{\mathrm{e}})$ is the Dirac delta function at the excitation location $x_{\mathrm{e}}$. The driving function $f_{\mathrm{e}}(t)$ resembles a pluck of a string and is of the following form \cite{Bilbao2019}:
\begin{equation}
f_{\mathrm{e}}(t)
=
\begin{cases}
\frac{1}{2}f_{\mathrm{amp}}
\big[
1
-
\cos
\big(
\frac{\pi t}{T_{\mathrm{e}}}
\big)
\big]
,
\quad
&t
\in
[
0,
T_{\mathrm{e}}
]
\\
0,
&\text{otherwise}
\end{cases}
\label{eq: excitation}
\end{equation}
Here $f_{\mathrm{amp}}$ is the excitation amplitude in \SI{}{N} and $T_{\mathrm{e}}$ is the excitation duration in \SI{}{s}. The excitation starting time is assumed to be zero.

%%%%%%%%%% EQUATION SCALING %%%%%%%%%%
\subsection{Equation Scaling}

In view of using the string model \eqref{eq: string} for dataset generation, it is useful to reduce the number of physical parameters to the smallest possible set. We employ spatial scaling by introducing normalised variables for position $\hat{x}=\frac{x}{L} \in [0,1]$ and displacement $\hat{u} = \frac{u}{L}$:
\begin{equation}
\begin{split}
\partial_{t}^2
\hat{u}
&=
\gamma^2
\partial_{\hat{x}}^2
\hat{u}
-
\kappa^2
\partial_{\hat{x}}^4
\hat{u}
-
2 \sigma_0
\partial_t
\hat{u}
+
2 \hat{\sigma}_1
\partial_t\partial_{\hat{x}}^2
\hat{u}
+{}\\
&{}+
\gamma^2
\frac{\alpha^2 - 1}{2}
\partial_{\hat{x}}
\big(
\mathcal{V}'(\xi)
\big)
+
\delta(\hat{x} - \hat{x}_{\mathrm{e}})
\hat{f}_{\mathrm{e}}(t),
\end{split}
\label{eq: string_scaled}
\end{equation}
where 
$\gamma = \frac{1}{L}\sqrt{\frac{T_0}{\rho A}}$,
$\kappa = \frac{1}{L^2}\sqrt{\frac{EI}{\rho A}}$,
$\alpha = \sqrt{\frac{EA}{T_0}}$,
$\hat{\sigma}_1 = \frac{\sigma_1}{L^2}$
and
$\hat{f}_{\mathrm{e}}(t) = \frac{1}{\rho A L^2} f_{\mathrm{e}}(t)$.

Thus, we have reduced a set of physical parameters $\{L, \rho, A,\allowbreak T_0, E, I, \sigma_0, \sigma_1\}$ to a set of only five parameters $\{\gamma, \kappa, \alpha, \sigma_0, \hat{\sigma}_1\}$. In the following sections we omit the ``hat'' while referring to the scaled string model \eqref{eq: string_scaled}, including the loss parameter $\hat{\sigma}_1$ and the excitation function $\hat{f}_{\mathrm{e}}(t)$.

%%%%%%%%%% MODAL DECOMPOSITION %%%%%%%%%%
\section{MODAL DECOMPOSITION}
\label{sec: modal}

The solution to equation \eqref{eq: string_scaled} can be decomposed into a set of modes \cite{BilbaoNSS}, yielding a finite-dimensional system when truncated to finite order $M$. The transverse displacement $u$ is rewritten as a superposition of modal displacements $\mathbf{q} = \mathbf{q}(t) = [q_{1}(t),\dots,q_{M}(t)]\tran$:
\begin{equation}
u(x, t)
=
\sum\limits_{m=1}^{M}
\Phi_m(x) q_m(t)
=
\boldsymbol{\Phi}\tran(x)
\mathbf{q}(t)
,
\label{eq: superposition}
\end{equation}
where modal shapes $\Phi_m(x) = \sqrt{2}\sin(m \pi x),\;m=1,\dots,M$ correspond to the solution of the eigenvalue problem for a stiff string under simply supported boundary conditions.

Substituting \eqref{eq: superposition} into \eqref{eq: string_scaled}, left-multiplying by $\bm{\Phi}(x)$ and taking an $L^2$ inner product over the interval $[0, 1]$, we obtain the following second-order system of ODEs:
\begin{equation}
\ddot{\mathbf{q}}
+
2 \bm{\Sigma}
\dot{\mathbf{q}}
+
\bm{\Omega}^2
\mathbf{q}
=
\nu^2
\mathbf{f}(\mathbf{q})
+
\bm{\Phi}(x_{\mathrm{e}})
f_{\mathrm{e}}(t),
\label{eq: modal_equation_second}
\end{equation}
where $\nu = \gamma \sqrt{\frac{\alpha^2 - 1}{2}}$ is treated as an independent parameter and matrices $\bm{\Sigma}$ and $\bm{\Omega}$ are defined using a $M \times M$ diagonal matrix $\mathbf{B}$ of modal wavenumbers $[\mathbf{B}]_{mm} = m\pi,\;m=1,\dots,M$ as:
\[
\bm{\Sigma}
=
\sigma_0
+
\sigma_1
\mathbf{B}^2
,
\quad
\bm{\Omega}^2
=
\gamma^2
\mathbf{B}^2
+
\kappa^2
\mathbf{B}^4
.
\]

The system \eqref{eq: modal_equation_second} can be rewritten in the first-order form using modal velocities $\mathbf{p} = \mathbf{p}(t) = [p_1(t),\dots,p_M(t)]\tran$ as:
\begin{equation}
\begin{cases}
\dot{\mathbf{q}}
=
\mathbf{p} 
\\
\dot{\mathbf{p}}
=
-
2 \bm{\Sigma}
\mathbf{p}
-
\bm{\Omega}^2
\mathbf{q}
+
\nu^2
\mathbf{f}(\mathbf{q})
+
\bm{\Phi}(x_{\mathrm{e}})
f_{\mathrm{e}}(t)
\end{cases}
\label{eq: modal_equation}
\end{equation}

%%%%%%%%%% SPECTRAL METHOD %%%%%%%%%%
\subsection{Spectral Method}
\label{sec: spectral_method}

To obtain a closed-form expression for the dimensionless nonlinearity $\mathbf{f}(\mathbf{q})\!\!: \mathbb{R}^M \rightarrow \mathbb{R}^M$, we employ a spectral method \cite{VichnevetskyBowlesFourierAnalysis, TrefethenSpectralMethods} to calculate spatial derivatives $\partial_x$ on the grid $x_{l+\sfrac{1}{2}} = \frac{1}{M+1}\big(l + \frac{1}{2}\big),\;l = 0,\dots,M$. Compared to our previous work \cite{Zheleznov2025} that calculated spatial derivatives $\partial_x$ over a continuous domain, the spectral method results in a significantly more efficient expression for $\mathbf{f}(\mathbf{q})$ and allows to implement the nonlinear function \eqref{eq: potential} without approximation by Taylor series.

First, we calculate spatial derivatives in a modal domain by multiplying modal displacements $\mathbf{q}$ by their respective wavenumbers $\mathbf{B}$. Second, we use the discrete cosine transform (DCT) to obtain spatial derivatives $\bm{\xi} = [\xi_{\sfrac{1}{2}},\dots,\xi_{M+\sfrac{1}{2}}]\tran$ on the grid $\{x_{l+\sfrac{1}{2}}\}_{l=0}^M$. Finally, we apply the elementwise nonlinear function $\mathcal{V}'(\bm{\xi}) =\allowbreak [\mathcal{V}'(\xi_{\sfrac{1}{2}}),\dots,\mathcal{V}'(\xi_{M+\sfrac{1}{2}})]\tran$, perform the inverse transform and another spatial differentiation in a modal domain. The closed-form expression for $\mathbf{f}(\mathbf{q})$ takes the form:
\[
\mathbf{f}(\mathbf{q})
=
-
\frac{1}{\sqrt{M+1}}
\mathbf{B}\mathbf{C}
\mathcal{V}'
(\bm{\xi}),
\quad
\bm{\xi}
\triangleq
\sqrt{M+1}
\mathbf{C}\tran
\mathbf{B}
\mathbf{q}
,
\]
where $\mathbf{C}$ is a truncated $M \times (M+1)$ matrix for the orthonormal DCT-II:
\[
[\mathbf{C}]_{ml}
=
\sqrt{
\frac{2}{M+1}
}
\cos
\bigg(
\frac{\pi}{M+1}
m
\bigg(
l + \frac{1}{2}
\bigg)
\bigg),
\quad
\substack{
m=1,\dots,M\\
l=0,\dots,M
}
.
\]

Using the fundamental theorem for line integrals \cite{HobsonMathMethods}, it is possible to derive a potential $V(\mathbf{q})$ for the dimensionless nonlinearity $\mathbf{f}(\mathbf{q})$: 
\begin{equation}
\mathbf{f}(\mathbf{q})
=
-\nabla_{\mathbf{q}} V(\mathbf{q}),
\quad
V(\mathbf{q})
=
\frac{1}{M+1}
\sum\limits_{l=0}^{M}
\mathcal{V}
\big(
\xi_{l+\sfrac{1}{2}}
\big)
.
\label{eq: potential_spectral}
\end{equation}

%%%%%%%%%% NUMERICAL SOLVER %%%%%%%%%%
\section{NUMERICAL SOLVER}
\label{sec: numerical_solver}

%%%%%%%%%% QUADRATISATION %%%%%%%%%%
\subsection{Quadratisation}
\label{sec: quadratisation}

In our previous work \cite{Zheleznov2025}, we employed the St{\"o}rmer-Verlet numerical solver which does not guarantee stability in simulation. To address this, we make use of the SAV technique \cite{Bilbao2023, Russo2024}. We introduce an auxiliary variable $\psi$ in order to quadratise the potential $V(\mathbf{q})$:
\[
\psi \triangleq \sqrt{2 V(\mathbf{q}) + \epsilon},
\]
where $\epsilon > 0$ is a gauge constant. The auxiliary variable $\psi$ is well defined if the potential $V(\mathbf{q})$ \eqref{eq: potential_spectral} is non-negative which is the case for the chosen nonlinear function \eqref{eq: potential}.

After quadratisation, we append the auxiliary variable $\psi$ to the state vector and rewrite the system \eqref{eq: modal_equation} as follows:
\begin{equation}
\begin{cases}
\dot{\mathbf{q}}
=
\mathbf{p} 
\\
\dot{\mathbf{p}}
=
-
2 \bm{\Sigma}
\mathbf{p}
-
\bm{\Omega}^2
\mathbf{q}
-
\nu^2
\mathbf{g}
\psi
+
\bm{\Phi}(x_{\mathrm{e}})
f_{\mathrm{e}}(t)
\\
\dot{\psi}
=
\mathbf{g}\tran
\mathbf{p}
\end{cases}
\label{eq: modal_equation_quadratisation}
\end{equation}
where $\mathbf{g} \triangleq \mathbf{g}_{\mathrm{std}}(\mathbf{q}) = \nabla_{\mathbf{q}}\psi$.

Furthermore, we add a control term to the system \eqref{eq: modal_equation_quadratisation} as described by Risse et al.\@ \cite{Risse2025} to reduce the drift between numerical values of $\psi$ and $\sqrt{2 V(\mathbf{q}) + \epsilon}$ in discrete time. This results in modification of the coupling term $\mathbf{g}$:
\begin{align*}
&\mathbf{g}
\triangleq
\mathbf{g}_{\mathrm{std}}(\mathbf{q})
+
\mathbf{g}_{\mathrm{mod}}(\mathbf{q}, \mathbf{p}, \psi)
, \\
&\mathbf{g}_{\mathrm{mod}}(\mathbf{q}, \mathbf{p}, \psi)
=
-\lambda_0
\Big(
\psi
-
\sqrt{2 V(\mathbf{q}) + \epsilon}
\Big)
\frac{\sign(\mathbf{p})}{\sign(\mathbf{p})\tran\mathbf{p}}
.
\end{align*}

%%%%%%%%%% TIME DISCRETISATION %%%%%%%%%%
\subsection{Time Discretisation}

We choose a time step $k$ in \SI{}{s}, yielding a sampling rate $f_{\mathrm{s}} = \frac{1}{k}$, and approximate state variables by the time series $\mathbf{q}^{n+\sfrac{1}{2}}$, $\mathbf{p}^n$ and $\psi^n$ on interleaved grids $t^{n+\sfrac{1}{2}} = \big(n+\frac{1}{2}\big)k$ and $t^n=nk$ for $n=0,\dots,N-1$. 
Using the difference operator $\delta_{t+}a^n = \frac{a^{n+1} - a^n}{k}$ and the averaging operator $\mu_{t+}a^n = \frac{a^{n+1} + a^n}{2}$, we define the scheme as \cite{Risse2025}:
\begin{equation}
\begin{cases}
\delta_{t+}
\mathbf{q}^{n-\sfrac{1}{2}}
=
\mathbf{p}^n
\\
\begin{split}
\delta_{t+}
\mathbf{p}^n
=
&-
2 \bm{\Sigma}
\mu_{t+}
\mathbf{p}^n
-
\bm{\Omega}^2
\mathbf{q}^{n + \sfrac{1}{2}}
-{}
\\
&{}-
\nu^2
\mathbf{g}^n
\mu_{t+}
\psi^n
+
\bm{\Phi}(x_{\mathrm{e}})
f_{\mathrm{e}}^{n + \sfrac{1}{2}}
\end{split}
\\
\delta_{t+}
\psi^n
=
(\mathbf{g}^n)\tran
\mu_{t+}
\mathbf{p}^n
\end{cases}
\label{eq: numerical_scheme}
\end{equation}
where
$
\mathbf{g}^n
=
\mathbf{g}_{\mathrm{std}}
\big(
\mathbf{q}^{n+\sfrac{1}{2}}
\big)
+
\mathbf{g}_{\mathrm{mod}}
\big(
\mu_{t+}\mathbf{q}^{n-\sfrac{1}{2}},
\mathbf{p}^n,
\psi^n
\big)
$
and
$f_{\mathrm{e}}^{n+\sfrac{1}{2}} = f_{\mathrm{e}}\big(t^{n+\sfrac{1}{2}}\big)$.

Numerical energy for the scheme \eqref{eq: numerical_scheme} is conserved and takes the form:
\[
E^n
=
\frac{1}{2}
(\mathbf{p}^n)\tran
\mathbf{p}^n
+
\frac{1}{2}
\big(
\mathbf{q}^{n+\sfrac{1}{2}}
\big)\tran
\bm{\Omega}^2
\mathbf{q}^{n-\sfrac{1}{2}}
+
\frac{\nu^2}{2}
(\psi^n)^2.
\]
The second term in $E^n$ can be bounded to guarantee non-negativity of numerical energy, resulting in the same stability condition $[\bm{\Omega}]_{MM} < \frac{2}{k}$ as for the linear system \cite{Bilbao2023}.

Defining $\mathbf{q}^n \triangleq \mu_{t+} \mathbf{q}^{n - \sfrac{1}{2}}$, the scheme \eqref{eq: numerical_scheme} can be written in an update form $(\mathbf{q}^n, \mathbf{p}^n, \psi^n) \rightarrow (\mathbf{q}^{n+1}, \mathbf{p}^{n+1}, \psi^{n+1})$:
\begin{equation}
\begin{cases}
\mathbf{q}^{n + \sfrac{1}{2}}
=
\mathbf{q}^n
+
\tfrac{k}{2}
\mathbf{p}^n
\\
\begin{split}
\mathbf{p}^{n+1}
&=
\Big[
\mathbf{I}
+
k\bm{\Sigma}
+
\tfrac{k^2\nu^2}{4}
\mathbf{g}^n
(\mathbf{g}^n)\tran
\Big]^{-1}
\times{}
\\
&{}\times
\Big[
\Big(
\mathbf{I}
-
k\bm{\Sigma}
-
\tfrac{k^2\nu^2}{4}
\mathbf{g}^n
(\mathbf{g}^n)\tran
\Big)
\mathbf{p}^n
+{}
\\
&{}+
k
\Big(
-\bm{\Omega}^2
\mathbf{q}^{n+\sfrac{1}{2}}
-
\nu^2
\mathbf{g}^n
\psi^n
+
\bm{\Phi}(x_{\mathrm{e}})
f_{\mathrm{e}}^{n + \sfrac{1}{2}}
\Big)
\Big]
\end{split}
\\
\mathbf{q}^{n+1}
=
\mathbf{q}^{n+\sfrac{1}{2}}
+
\tfrac{k}{2}
\mathbf{p}^{n+1}
\\
\psi^{n+1}
=
\psi^n
+
k
(\mathbf{g}^n)\tran
\tfrac{
\mathbf{p}^{n+1}
+
\mathbf{p}^n
}{2}
\end{cases}
\label{eq: numerical_scheme_update}
\end{equation}
The inverse appearing in \eqref{eq: numerical_scheme_update} can be easily computed with the Sherman-Morrison formula \cite{Sherman1950}. Using \eqref{eq: superposition}, we obtain an audio output as $w^n = \bm{\Phi}\tran(x_{\mathrm{o}}) \mathbf{q}^n$.

%%%%%%%%%% DIFFERENTIABLE MODEL %%%%%%%%%%
\section{DIFFERENTIABLE MODEL}
\label{sec: model}

%%%%%%%%%% NEURAL ODES %%%%%%%%%%
\subsection{Neural Ordinary Differential Equations}

NODEs can be defined through the following first-order system:
\begin{equation}
\frac{d\mathbf{y}}{dt}
=
\mathbf{h}_{\theta}
(\mathbf{y}, t)
,
\quad
\mathbf{y}(0)
=
\mathbf{y}_0.
\label{eq: neural_ode}
\end{equation}
Here $\mathbf{y} = \mathbf{y}(t)\!\!: \mathbb{R}^+ \rightarrow \mathbb{R}^K$ is an unknown function of time $t$, $\mathbf{y}_0 \in \mathbb{R}^K$ is an initial condition and $\mathbf{h}_{\theta}(\mathbf{y}, t)\!\!: \mathbb{R}^K \times \mathbb{R}^+ \rightarrow \mathbb{R}^K$ is a neural network where $\theta$ denotes the set of all learnable parameters and $K$ denotes the state dimension. Generally, a simple architecture such as an MLP is chosen for $\mathbf{h}_{\theta}(\mathbf{y}, t)$. Chen et al.\@ \cite{Chen2018} have showed that the system \eqref{eq: neural_ode} in combination with a numerical solver, labelled as ODENet, can be trained from the observed state data to reproduce the dynamics of a target system for which the theoretical model may be unknown.

Assume a target trajectory $\{\mathbf{y}_0,\mathbf{y}^1,\dots,\mathbf{y}^{N-1}\}$ sampled on the time grid $\{t^n\}_{n=0}^{N-1}$. Given a predicted trajectory $\{\mathbf{y}_0,\tilde{\mathbf{y}}^1,\dots,\tilde{\mathbf{y}}^{N-1}\}$ by a numerical solution to the initial value problem \eqref{eq: neural_ode}, i.e., a forward pass of the ODENet, we can construct an objective function $J(\theta)$ such as mean squared error (MSE):
\begin{equation}
J(\theta)
=
\MSE(\tilde{\mathbf{y}}^n, \mathbf{y}^n)
\triangleq
\frac{1}{KN}
\sum_{n=0}^{N-1}
\norm{
\tilde{\mathbf{y}}^n
-
\mathbf{y}^n
}_2^2
,
\label{eq: mse}
\end{equation}
where $\norm{\cdot}_2$ is the Euclidean norm. We search for a local minimum of $J(\theta)$ using gradient-based optimisation techniques where the gradient $\nabla_{\theta}J$ can be computed using the backpropagation algorithm \cite{Rumelhart1986} through internal operations of a numerical solver or the adjoint sensitivity method \cite{Pontryagin1962, Chen2018}. In most cases the objective function $J(\theta)$ will be averaged for a finite set of target trajectories before each optimisation step.

%%%%%%%%%% EXTENSION %%%%%%%%%%
\subsection{Extension for Modal Synthesis}

In the case of modal synthesis, there is a known ODE structure \eqref{eq: modal_equation} which can serve an inductive bias for a NODEs framework \cite{KidgerPhD, Lai2022}. In particular, we parametrise only a dimensionless memoryless nonlinear function $\mathbf{f}_{\theta}(\mathbf{q})\!\!: \mathbb{R}^M \rightarrow \mathbb{R}^M$ with a neural network, yielding a system of physics-informed NODEs:
\begin{equation}
\begin{bmatrix}
\dot{\mathbf{q}} \\
\dot{\mathbf{p}}
\end{bmatrix}
=
\underbrace{
\begin{bmatrix}
\hphantom{-}
\mathbf{0}
\hphantom{^2}
&
\hphantom{-}
\mathbf{I}
\hphantom{\bm{\Sigma}}
\\
-\bm{\Omega}^2
&
-2\bm{\Sigma}
\end{bmatrix}
\begin{bmatrix}
\mathbf{q} \\
\mathbf{p}
\end{bmatrix}
}_{\text{\makebox[0pt]{\scriptsize Linear vibration}}}
{}+{}
\nu^2
\underbrace{
\begin{bmatrix}
\mathbf{0} \\
\mathbf{f}_{\theta}(\mathbf{q})
\end{bmatrix}
}_{\text{\makebox[0pt]{\scriptsize Neural network}}}
+
\underbrace{
\begin{bmatrix}
\mathbf{0} \\
\bm{\Phi}(x_{\mathrm{e}})
\end{bmatrix}
f_{\mathrm{e}}(t)
}_{\text{\makebox[0pt]{\scriptsize Excitation}}}
\label{eq: modal_equation_neural}
\end{equation}
As mentioned earlier, initial conditions for a state vector $\mathbf{y}\tran = [\mathbf{q}\tran, \mathbf{p}\tran]$ are assumed to be zero. To compute a forward pass of the physics-informed ODENet, we:
\begin{itemize}
\itemsep0.5em

\item set $\bm{\Sigma}, \bm{\Omega}, \nu, \bm{\Phi}(x_{\mathrm{e}})$ in \eqref{eq: modal_equation_neural} using physical parameters of a target solution;

\item precompute $f_{\mathrm{e}}^{n+\sfrac{1}{2}} = f_{\mathrm{e}}\big(t^{n+\sfrac{1}{2}}\big),\;n=0,\dots,N-1$ using \eqref{eq: excitation} and excitation parameters of a target solution;

\item use the numerical solver \eqref{eq: numerical_scheme_update} as in the case of a regular system  to produce a predicted trajectory.

\end{itemize}

The formulation \eqref{eq: modal_equation_neural} has strong implications. The exact expression for linear vibration exploits the periodic, harmonic and lossy nature of a musical system. Thus, we aid the optimisation process by constraining the space of possible solutions and improve interpretability of the model. Furthermore, the neural network $\mathbf{f}_{\theta}(\mathbf{q})$ is memoryless and dimensionless, thus does not depend on physical parameters of a system and external excitation. Theoretically, these parameters can be changed after the training as long as range of displacements $\mathbf{q}$ stays the same as in a training dataset to simulate other configurations of a system. Some limitations exist since the neural network $\mathbf{f}_{\theta}(\mathbf{q})$ implicitly depends on boundary conditions and the number of modes.

%%%%%%%%%% GRADIENT NETWORKS %%%%%%%%%%
\subsection{Gradient Networks}

The nature of the SAV technique (Section \ref{sec: quadratisation}) constrains possible network architectures for $\mathbf{f}_{\theta}(\mathbf{q})$ as we require existence of a closed-form and non-negative potential $V_{\theta}(\mathbf{q})$. For example, we can no longer use MLPs as in our previous work \cite{Zheleznov2025}. To overcome this, we employ GradNets \cite{Chaudhari2025} that directly parametrise gradients of various function classes. In particular, we use the following architecture:
\begin{equation}
\mathbf{f}_{\theta}
(\mathbf{q})
=
-
\mathbf{W}\tran
[
\bm{\alpha}
\odot
\sigma
(
\mathbf{z}
)
]
,
\quad
\mathbf{z}
=
\bm{\beta}
\odot
\mathbf{W}
\mathbf{q}
+
\mathbf{b},
\label{eq: grad_net}
\end{equation}
where $\odot$ denotes the Hadamard product. Learnable parameters are the $H \times M$ weight matrix $\mathbf{W}$, bias $\mathbf{b} \in \mathbb{R}^H$ and scaling vectors $\bm{\alpha} \in \mathbb{R}^H$ and $\bm{\beta} \in \mathbb{R}^H$ where $H$ is the hidden dimension. An activation function $\sigma(x)\!\!:\mathbb{R} \rightarrow \mathbb{R}$ is applied elementwise as $\sigma(\mathbf{z}) = [\sigma(z_1),\dots,\sigma(z_H)]\tran$.

We can notice a strong resemblance between the GradNet \eqref{eq: grad_net} and the spectral method (Section \ref{sec: spectral_method}). The weight matrix $\mathbf{W}$ acts as a transform to a hidden space where we apply the elementwise nonlinear function $\sigma(\mathbf{z})$ before performing an ``inverse'' transform with the transposed weight matrix $\mathbf{W}\tran$. Moreover, the potential \eqref{eq: potential_spectral} arising from the spectral method is expressed as a sum of convex ridge functions and the GradNet \eqref{eq: grad_net} with a monotonically-increasing activation function $\sigma$ and non-negative scaling vectors $\bm{\alpha}$ and $\bm{\beta}$ can universally approximate gradients of such functions \cite{Chaudhari2025}.

Assuming existence of an antiderivative function $\phi(x)\!\!:\mathbb{R} \rightarrow \mathbb{R}$ so that $\sigma = \phi'$, the closed-form expression for $V_{\theta}(\mathbf{q})$ takes the form:
\[
\mathbf{f}_{\theta}
(\mathbf{q})
=
-\nabla_{\mathbf{q}}
V_{\theta}
(\mathbf{q}),
\quad
V_{\theta}
(\mathbf{q})
=
\sum_{i=1}^H
\frac{\alpha_i}{\beta_i}
\phi(z_i)
.
\]
If $\phi$ is a non-negative function, the potential $V_{\theta}(\mathbf{q})$ is also non-negative. The scaling vectors can be redefined to avoid ill-conditioning from division.

%%%%%%%%%% EVALUATION %%%%%%%%%%
\section{EVALUATION}
\label{sec: evaluation}

The physics-informed ODENet (Section \ref{sec: model}) was implemented in the PyTorch framework \cite{Paszke2019} and evaluated in the case of nonlinear transverse string vibration. The training was conducted on a cloud server equipped with NVIDIA GeForce RTX 2080 Ti GPUs with \SI{12}{GB} VRAM. The source code used for dataset generation and training is available in the accompanying GitHub repository\footnote{\url{https://github.com/victorzheleznov/jaes-modal-node}}.

For the parametrisation of $\mathbf{f}_{\theta}(\mathbf{q})$ we used the GradNet \eqref{eq: grad_net} with hidden dimension $H = 1000$. This number was chosen based on available GPU memory since the validation loss was found to consistently decrease with increasing hidden dimension. However, smaller networks may also be sufficient from the perceptual point of view. A leaky rectified linear unit was used as an activation $\sigma$ which is a monotonically-increasing function with a non-negative antiderivative $\phi$. Kaiming initialisation \cite{He2015} was used for the initial weights $\mathbf{W}$ and the biases $\mathbf{b}$ were initialised to zero. Logarithms of scaling vectors $\bm{\alpha}$ and $\bm{\beta}$ were used as learnable parameters to enforce non-negativity and were initialised around zero using a normal distribution.

The training loss was the MSE \eqref{eq: mse} taken over the state vector $\mathbf{y}\tran = [\mathbf{q}\tran, \mathbf{p}\tran]$, excluding the auxiliary variable $\psi$ arising from the SAV method. This implied that the potential $V_{\theta}(\mathbf{q})$ could be shifted by a constant relative to the target potential \eqref{eq: potential_spectral} since dynamics of the system and the training loss would not be affected. Backpropagation was performed using internal operations of the numerical solver \eqref{eq: numerical_scheme_update}, i.e., the ``discretise-then-optimise'' method, which is generally a preferred approach due to its gradient accuracy, speed and straightforward implementation \cite{KidgerPhD}. Calculation of the control term $\mathbf{g}_{\mathrm{mod}}$ was excluded from the computational graph. In some cases, it was found to significantly compromise training by steering the potential $V_{\theta}(\mathbf{q})$ to incorrect values of the auxiliary variable $\psi$ and lead to an unexpected increase in the training loss. As only gradient calculation is affected, stability and drift regulation properties of the numerical solver \eqref{eq: numerical_scheme_update} still hold during inference. The Adam optimiser \cite{Kingma2017} was used with default parameters. The training was performed for 2000 epochs using a cluster of four GPUs to parallelise batches of data. The resulting model was chosen based on the lowest validation loss obtained during optimisation.

For training we used a variation of the teacher forcing technique \cite{GoodfellowDL} by splitting up a target trajectory into \SI{1}{ms} segments and providing true initial conditions for each segment to the ODENet. Since the numerical solver \eqref{eq: numerical_scheme_update} is given as a one step update we had access to both displacement and velocity of the target numerical solution at each time step, and thus initial conditions for each segment. The auxiliary variable was initialised as $\psi_0 = \sqrt{2 V_{\theta}(\mathbf{q}_0) + \epsilon}$ using initial displacements $\mathbf{q}_0$ to produce a consistent initial condition \cite{Risse2025}. In addition, the excitation function \eqref{eq: excitation} was shifted in time to reflect a new starting point for integration. The main reason for using this technique was to speed up training as the number of integration steps in the numerical solver is significantly reduced. These integration steps cannot be parallelised in time. Moreover, the likelihood of vanishing and exploding gradient problems during optimisation is also reduced by this technique as backpropagation on long time series is avoided \cite{KidgerPhD}.

%%%%%%%%%% DATASETS %%%%%%%%%%
\subsection{Datasets}
\label{sec: datasets}

Three separate datasets for training, validation and testing were independently generated using the nonlinear transverse string model (Section \ref{sec: string}). The training and test datasets both consisted of 60 trajectories and the validation dataset consisted of 20 trajectories. Each trajectory included both displacement and velocity information for each mode of a string. The number of modes was equal to $M = 75$. Taking into account the effect of stiffness, this covered the \SI{10}{kHz} range for the lowest considered fundamental frequency at \SI{61.74}{Hz}. Oversampling by a factor of two for standard sampling rates \SI{44.1}{kHz} and \SI{48}{kHz} was used to avoid aliasing due to the nonlinear effects for high amplitude excitations.

\begin{table}[h]
\def\arraystretch{1.1}
\centering
\caption{Simulation parameters used for datasets generation.}
\label{tab: datasets}
\begin{tabular}{lll}
\hline
Parameter
&
Training
&
Validation and Test
\\
\hline
$f_{\mathrm{s}}$
&
\SI{88.2}{kHz}
&
\SI{96}{kHz}
\\
$T_{\mathrm{sim}}$
&
\SI{2}{s}
&
\SI{3}{s}
\\
$T_{\mathrm{e}}$
&
(\num{0.5} to \num{1.5}) \SI{}{ms}
& 
(\num{0.5} to \num{1.5}) \SI{}{ms}
\\
$f_{\mathrm{amp}}$
& 
(\num{2.5} to \num{3.5})\ensuremath{{}\times{}}\num{e4}
& 
(\num{3.5} to \num{5})\ensuremath{{}\times{}}\num{e4}
\\
$\gamma$
&
\num{123.48} to \num{174.62}
&
\num{174.62} to \num{246.94}
\\
$\kappa$
&
\num{1.01} to \num{1.05}
&
\num{1.05} to \num{1.1}
\\
$\nu$
&
\num{123.48} to \num{174.62}
&
\num{123.48} to \num{174.62}
\\
$\sigma_0$
&
$3$
&
$2$
\\
$\sigma_1$
&
\num{2e-4}
&
\num{2e-4}
\\
$x_{\mathrm{e}}$
&
\num{0.1} to \num{0.9}
&
\num{0.1} to \num{0.9}
\\
$x_{\mathrm{o}}$
&
\num{0.1} to \num{0.9}
&
\num{0.1} to \num{0.9}
\\
\hline
\end{tabular}
\end{table}

Simulation parameters used for the datasets are provided in Table \ref{tab: datasets}, where $T_{\mathrm{sim}}$ corresponds to the duration of simulation. For specified parameter ranges, randomised values were generated from a uniform distribution. For training, fundamental frequencies, equal to $\frac{\gamma}{2}$, spanned a half-octave range from \SI{61.74}{Hz} (B1 note) to \SI{87.31}{Hz} (F2 note). For validation and testing, the next half-octave range from \SI{87.31}{Hz} (F2 note) to \SI{123.47}{Hz} (B2 note) was used. In addition, the simulation duration was increased to account for a longer decay time due to a smaller loss parameter $\sigma_0$. Non-overlapping ranges for the stiffness parameter $\kappa$ were used. Strings were excited by randomly-generated excitation functions \eqref{eq: excitation} at randomised excitation positions $x_{\mathrm{e}}$. The range for the excitation amplitude $f_{\mathrm{amp}}$ was scaled linearly with fundamental frequencies to preserve the strength of the nonlinear effects between the datasets, and thus the range of displacements $\mathbf{q}$ (see Appendix \ref{sec: excitation_amplitude}). An audio output was drawn from randomised positions $x_{\mathrm{o}}$ along a string for each trajectory.

These simulation parameters were motivated by two considerations. First, we wanted to test generalisation of the model to physical parameters, sampling rates and time scales not seen during training. In view of other machine learning approaches, this flexibility and controllability of the physics-informed ODENet can be considered as its main advantage. Second, strings with low fundamental frequencies were chosen so that the nonlinear effects were more prominent in simulations \cite{BilbaoNSS}. Since the model architecture is designed to learn the residual between the linear and nonlinear solutions, the datasets needed to reflect a significant difference between them.

%%%%%%%%%% RESULTS %%%%%%%%%%
\subsection{Results}

\begin{figure}[t]
\center
\includegraphics[width=\columnwidth]{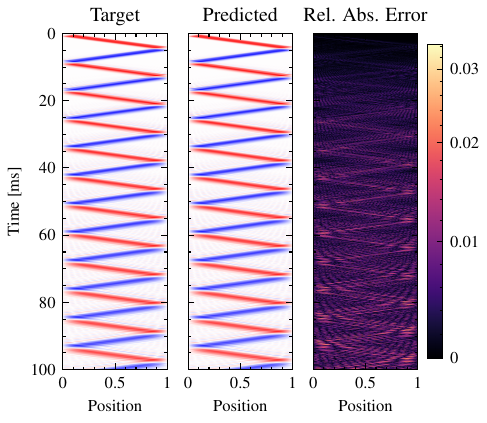}
\caption{The predicted displacement trajectory with the largest $\MSE_{\mathrm{rel}}(\tilde{w}^n, w^n)$ (centre), the target displacement trajectory (left) and the pointwise absolute error between them relative to the maximum absolute value of target trajectory (right).}
\label{fig: 40_displacement_grid}
\end{figure}

\begin{figure}[t]
\center
\includegraphics[width=\columnwidth]{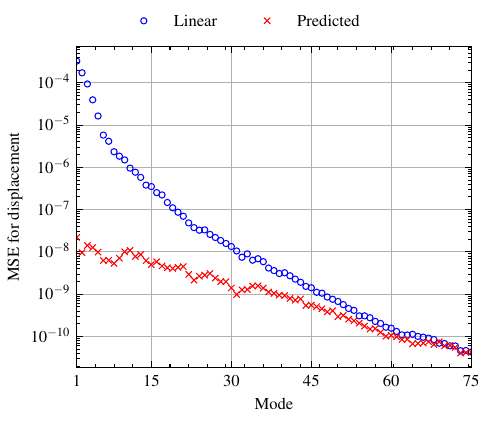}
\caption{MSE per mode for the initial \SI{100}{ms} of the predicted and linear displacement trajectories compared to the target solution.}
\label{fig: mse_q_per_mode_slice_test}
\end{figure}

\begin{figure*}[p]
\center
\includegraphics[width=\textwidth]{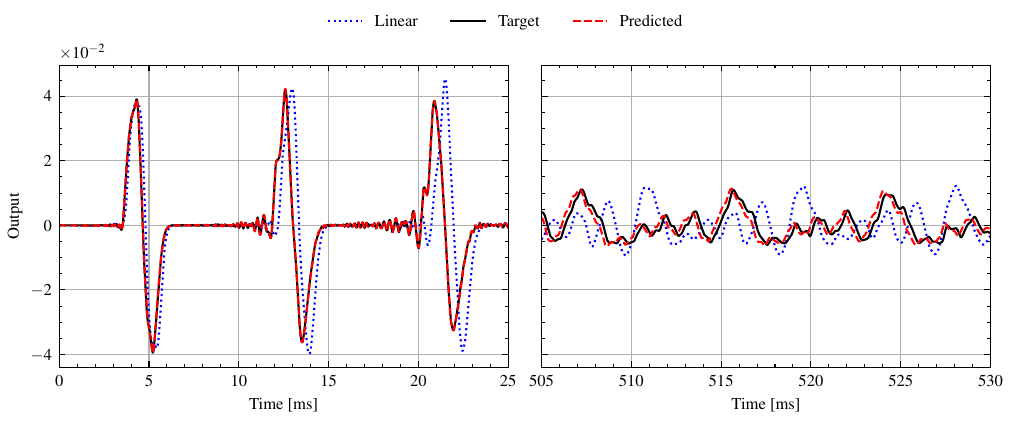}
\caption{The predicted audio output with the largest $\MSE_{\mathrm{rel}}(\tilde{w}^n,w^n)$. Taken at normalised position $x_\mathrm{o} = 0.89$.}
\label{fig: 40_wave}
\end{figure*}

\begin{figure*}[p]
\center
\includegraphics[width=\textwidth]{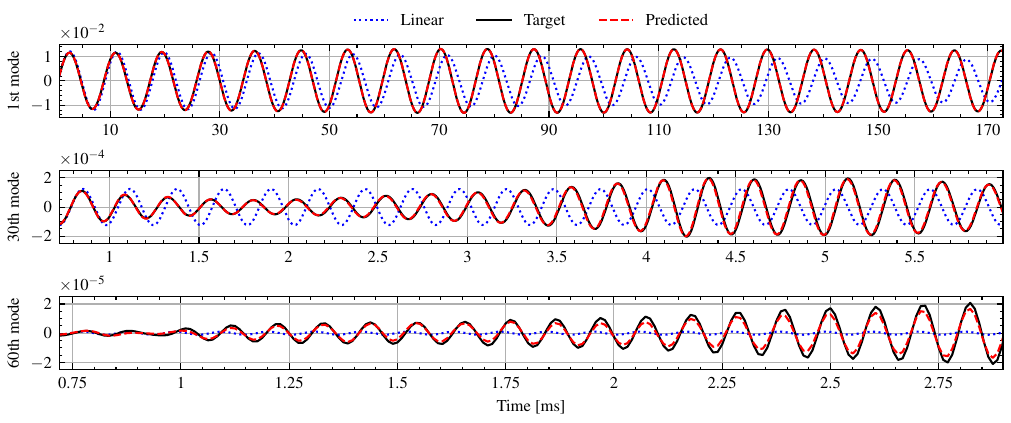}
\caption{Displacements of the 1st, 30th and 60th mode for the predicted trajectory with the largest $\MSE_{\mathrm{rel}}(\tilde{w}^n,w^n)$. Taken at normalised position $x_\mathrm{o} = 0.89$. Initial 20 periods after the excitation are shown.}
\label{fig: 40_displacement}
\end{figure*}

\begin{figure*}[p]
\center
\includegraphics[width=\textwidth]{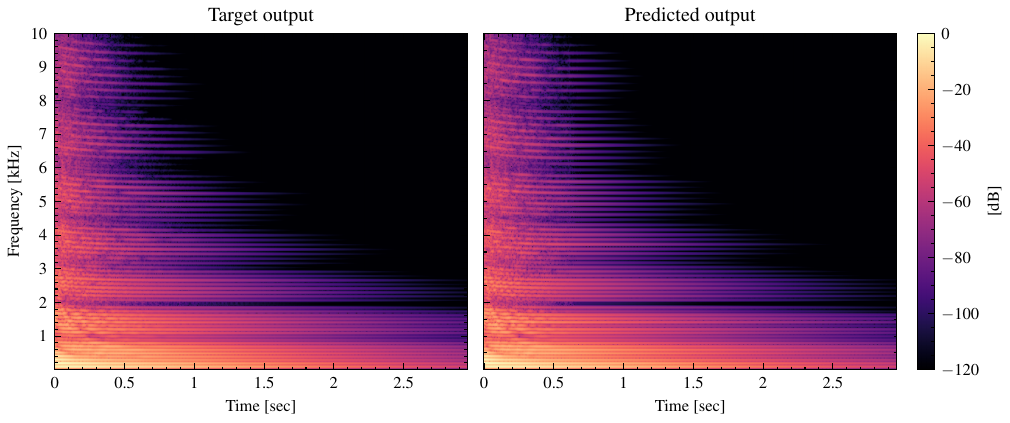}
\caption{Spectrogram of the predicted audio output with the strongest nonlinear effects. Taken at normalised position $x_\mathrm{o} = 0.89$.}
\label{fig: 6_spec}
\end{figure*}

For evaluation of the trained model, we used the relative MSE and the relative mean absolute error (MAE), defined for a general time series $\mathbf{x}^n$ as:
\begin{align*}
&\MSE_{\mathrm{rel}}
(\tilde{\mathbf{x}}^n, \mathbf{x}^n)
\triangleq
\frac{
\sum_n
\norm{\tilde{\mathbf{x}}^n - \mathbf{x}^n}_2^2
}
{
\sum_n
\norm{\mathbf{x}^n}_2^2
},
\\
&\MAE_{\mathrm{rel}}
(\tilde{\mathbf{x}}^n, \mathbf{x}^n)
\triangleq
\frac{
\sum_n
\norm{\tilde{\mathbf{x}}^n - \mathbf{x}^n}_1
}
{
\sum_n
\norm{\mathbf{x}^n}_1
},
\end{align*}
where $\tilde{\mathbf{x}}^n$ is a prediction given by the model.

Metrics for predicted displacement trajectories $\tilde{\mathbf{q}}^n$ and audio outputs $\tilde{w}^n$, averaged over the training, validation and test datasets, are provided in Table \ref{tab: metrics}. Metrics were separately computed for the initial \SI{100}{ms} and for the full duration of simulation. As can be seen, metrics for the validation and test datasets have the same order of magnitude as for the training dataset and do not degrade for unseen simulation parameters. This suggests that the sampling rate and duration of simulation can be changed after the training and the model generalises to unseen physical parameters as long as the range of displacements $\mathbf{q}^n$ stays the same. In addition, values of the relative MAE are close to the square roots of the relative MSE, indicating the absence of outliers in the datasets.

\begin{table}[h]
\def\arraystretch{1.1}
\addtolength{\tabcolsep}{-0.2em}
\centering
\caption{Metrics for the nonlinear transverse string experiment.}
\label{tab: metrics}
\begin{tabular}{lccc}
\hline
Metric & Training & Validation & Test
\\
\hline
\multicolumn{4}{c}{
Computed for initial \SI{100}{ms}
}
\\
$\MSE_{\mathrm{rel}}
(\tilde{\mathbf{q}}^n,\mathbf{q}^n)$
&
\num{2.8e-4}
&
\num{2.0e-4}
&
\num{2.7e-4}
\\
$\MSE_{\mathrm{rel}}
(\tilde{w}^n,w^n)$
&
\num{3.3e-4}
&
\num{1.7e-4}
&
\num{2.7e-4}
\\
$\MAE_{\mathrm{rel}}
(\tilde{\mathbf{q}}^n,\mathbf{q}^n)$
&
\num{4.1e-2}
&
\num{3.4e-2}
&
\num{3.4e-2}
\\
$\MAE_{\mathrm{rel}}
(\tilde{w}^n,w^n)$
&
\num{1.6e-2}
&
\num{1.1e-2}
&
\num{1.3e-2}
\\
\multicolumn{4}{c}{
Computed for full duration
}
\\
$\MSE_{\mathrm{rel}}
(\tilde{\mathbf{q}}^n,\mathbf{q}^n)$
&
\num{5.4e-2}
&
\num{7.0e-2}
&
\num{6.9e-2}
\\
$\MSE_{\mathrm{rel}}
(\tilde{w}^n,w^n)$
&
\num{5.5e-2}
&
\num{6.6e-2}
&
\num{7.3e-2}
\\
$\MAE_{\mathrm{rel}}
(\tilde{\mathbf{q}}^n,\mathbf{q}^n)$
&
\num{3.6e-1}
&
\num{3.9e-1}
&
\num{3.9e-1}
\\
$\MAE_{\mathrm{rel}}
(\tilde{w}^n,w^n)$
&
\num{3.1e-1}
&
\num{3.3e-1}
&
\num{3.5e-1}
\\
\hline
\end{tabular}
\end{table}

To illustrate a worst-case example, we selected a predicted trajectory from the test dataset with the largest $\MSE_{\mathrm{rel}}(\tilde{w}^n, w^n)$ considering the full duration of simulation. This corresponded to a string with a \SI{116.25}{Hz} fundamental frequency. As can be seen on Fig.\@ \ref{fig: 40_displacement_grid}, the predicted displacement trajectory maintains the structure of the target solution but the error between them grows over time. This is expected, as any difference between the learned $\mathbf{f}_{\theta}(\mathbf{q})$ and the underlying $\mathbf{f}(\mathbf{q})$ nonlinearity is gradually accumulated by a numerical solver. This is also confirmed by the metrics in Table \ref{tab: metrics} which rise when evaluated for the full duration of simulation compared to the initial \SI{100}{ms}. However, as the nonlinear effects become less prominent over time due to loss in the system \cite{BilbaoNSS}, the initial response of the model to an external excitation is significantly more important for capturing the nonlinear behaviour.

Looking at the audio output in Fig.\@ \ref{fig: 40_wave}, we see that initially the predicted waveform closely follows the target solution, including high-frequency partials of higher modes. Moving forward in time, the predicted waveform still preserves the shape of the target solution. The predicted displacements for individual modes also closely follow the target solution (Fig.\@ \ref{fig: 40_displacement}). In addition, we see that the linear solution is significantly different, meaning the network $\mathbf{f}_{\theta}(\mathbf{q})$ has learned a significant nonlinear effect in addition to the linear vibration that is treated as known in the model. This is especially evident for the 60th mode in Fig.\@ \ref{fig: 40_displacement}, where the energy transfer between the modes due to the nonlinear coupling leads to a significant increase in amplitude, which is reproduced by the model.

We can further analyse the model by examining the MSE of the predicted and linear displacement trajectories compared to the target solution for each mode individually. Fig.\@ \ref{fig: mse_q_per_mode_slice_test} shows these metrics which were computed over the whole test dataset for the initial \SI{100}{ms} of simulation. We see that the predicted displacement trajectories are much closer to the target solution, with up to four orders of magnitude smaller error than the linear solution. Since absolute values of displacements are significantly smaller for higher modes, the training loss is dominated by the lower modes and the network $\mathbf{f}_{\theta}(\mathbf{q})$ does not capture the higher modes as accurately as the lower modes. This leads to a faster error accumulation for higher modes due to integration by a numerical solver. However, if we consider a trajectory from the test dataset with the strongest nonlinear effects (\SI{98.18}{Hz} fundamental frequency), we can see on the spectrogram in Fig.\@ \ref{fig: 6_spec} that the pitch glide effect is reproduced for all of the modes in line with the target solution. This suggests that errors for higher modes are mainly caused by incorrectly estimated amplitudes rather than instantaneous frequencies.

Formal perceptual tests are out of scope of this paper and readers are encouraged to listen to sound examples presented on the accompanying page\footnote{\url{https://victorzheleznov.github.io/jaes-modal-node}}. Based on informal listening, the predicted and target audio were found to be nearly indistinguishable, while the difference compared to the linear baseline was clearly audible.

%%%%%%%%%% CONCLUSION %%%%%%%%%%
\section{CONCLUSION}

We have presented here a differentiable modal synthesis model that is capable of learning nonlinear dynamics from data. The proposed approach separates the linear vibration of a system from the nonlinear coupling between the modes described by a dimensionless potential function. We employed gradient networks (GradNets) \cite{Chaudhari2025} that universally approximate gradients of such functions and made use of the scalar auxiliary variable (SAV) technique \cite{Bilbao2023, Russo2024, Risse2025} to ensure numerical stability in simulation. For the case study of nonlinear transverse string vibration, we have shown that the sampling rate and duration of simulation can be easily changed after the training and the model generalises to unseen physical parameters, resulting in a flexible and controllable sound synthesis approach.

Future work will be focused on using this framework for learning from acoustic recordings of string instruments. This will introduce additional challenges. First, modal frequencies and the loss profile must be estimated from audio spectrograms because physical parameters of a string will be unknown. Second, external excitation should be formulated as an initial condition (e.g., a triangular function) since estimation of an excitation force would require additional sensors \cite{Mehes2016}. Finally, the training methodology must be robust to noise and based on audio waveforms as we will not have the reference data for displacements and velocities of each mode. It is theorised that the proposed approach could be used to expand the timbral range of a real instrument in a digital domain and synthesise sounds which were not part of the recorded data by changing physical parameters after the training. In addition, physics-informed architectures like that presented here have the potential to address problems where the underlying physical phenomena are not fully understood, e.g., the bowed string in musical acoustics \cite{Galluzzo2017}.

%%%%%%%%%% BIBLIOGRAPHY %%%%%%%%%%
\printbibliography

%%%%%%%%%% APPENDIX %%%%%%%%%%
\appendix
\renewcommand\thesection{A}
\section*{APPENDIX}

\subsection{Adjustment for Excitation Amplitude}
\label{sec: excitation_amplitude}

We consider the following initial value problem for a harmonic oscillator under the driving function $f_{\mathrm{e}}(t)$ \eqref{eq: excitation}:
\[
\ddot{q}
+
\omega^2
q
=
f_{\mathrm{e}}(t),
\quad
q(0) = 0,
\quad
\dot{q}(0) = 0.
\]
Using the Laplace transform, we can obtain terms of the solution oscillating at the frequency $\omega$:
\begin{align*}
q(t)
\approx
&-
f_{\mathrm{amp}}
\frac
{
\big(
\frac{\omega_{\mathrm{e}}}{2}\big
)^2
}
{
2
\big[
\big(
\frac{\omega_{\mathrm{e}}}{2}\big
)^2
-
\omega^2
\big]
\omega^2
}
\cos(\omega t)
+{}
\\
&{}+
f_{\mathrm{amp}}
\frac
{
\big(
\frac{\omega_{\mathrm{e}}}{2}\big
)^2
-
2\omega^2
}
{
2
\big[
\big(
\frac{\omega_{\mathrm{e}}}{2}\big
)^2
-
\omega^2
\big]
\omega^2
}
\cos(\omega(t - T_{\mathrm{e}})),
\end{align*}
where $\omega_{\mathrm{e}} \triangleq \frac{2\pi}{T_{\mathrm{e}}}$.
Under the assumption $\omega_{\mathrm{e}} \gg \omega$, we can approximate $\cos(\omega(t - T_{\mathrm{e}})) \approx \cos(\omega t) - \omega T_{\mathrm{e}}\sin(\omega t)$:
\begin{align*}
q(t)
\approx
&-
f_{\mathrm{amp}}
\frac
{1}
{
\big(
\frac{\omega_{\mathrm{e}}}{2}\big
)^2
-
\omega^2
}
\cos(\omega t)
-{} \\
&{}-
f_{\mathrm{amp}}
\frac
{
\big[
\big(
\frac{\omega_{\mathrm{e}}}{2}\big
)^2
-
2\omega^2
\big]
T_{\mathrm{e}}
}
{
2
\big[
\big(
\frac{\omega_{\mathrm{e}}}{2}\big
)^2
-
\omega^2
\big]
\omega
}
\sin(\omega t).
\end{align*}
Amplitude of the displacement $q(t)$ is dominated by the second term and can be approximated as $f_{\mathrm{amp}}\frac{T_{\mathrm{e}}}{2\omega}$. Thus, we can keep the range of displacement $q(t)$ roughly the same by linearly scaling the excitation amplitude $f_{\mathrm{amp}}$ with the frequency $\omega$.

\end{document}